\documentstyle[preprint,aps]{revtex}
\begin{document}
\draft
\title{\bf{\LARGE{Aharonov--Bohm effect in the presence of
evanescent modes}}}
\author{B. C. Gupta, P. Singha Deo and A. M. Jayannavar}
\address{Institute of physics, Bhubaneswar - 751 005, India}
\maketitle
\vskip 2.5cm
\begin{abstract}
{\small It is known that differential magnetoconductance of a
normal metal loop connected to reservoirs by ideal wires is
always negative when an electron travels as an evanescent modes
in the loop. This is in contrast to the fact that the
magnetoconductance for propagating modes is very sensitive to
small changes in geometric details and the Fermi energy and
moreover it can be positive as well as negative. Here we explore
the role of impurities in the leads in determining the
magnetoconductance of the loop.  We find that the change in
magnetoconductance is negative and can be made large provided
the impurities do not create resonant states in the systems.
This theoretical finding may play an useful role in quantum
switch operations.}
\end{abstract}
%\narrowtext
%\thispagestyle{empty}
\newpage
\noindent{\bf Introduction}

\noindent Recent advances in microfabrication technology
\cite{kirk,been1,kram,alts1} have made it possible to 
fabricate artificial structures with very fine control over
design parameters and one has approached a technological level
of solid state structures in which the energy and length scales
are such that macroscopic quantum effects can be observed.
Typical sizes of these systems vary from nanometer to
micrometer.  At very low temperatures (typically mK), inelastic
scattering is significantly suppressed and the electron phase
coherence length can become large compared to the system size.
In this regime the electron maintains the single particle phase
coherence across the entire sample and the system is called a
mesoscopic system \cite{kirk,been1,kram,alts1}.  The idealized
sample becomes an electron wave guide and transport properties
are solely determined by the impurity configuration and geometry
of the conductor and by the principles of quantum mechanics.
Transport in these systems cannot be described by standard
classical Boltzmannian theory, where self averaging over many
macroscopic configurations is assumed. Instead, a mesoscopic
system as a whole should be treated as a phase coherent elastic
scatterer.

It has now become clear that transport behavior in these
nanostructures (in the regime of quantum ballistic transport),
resemble in many ways, properties of wave guides in
electromagnetic wave propagation. This similarity with guided
electromagnetic wave propagation has opened up the possibility
of new quantum semiconductor devices \cite{bout}.  These quantum
devices rely on quantum effects for their operation, and are
based on interferometric principles. The mechanism of switch
operation by quantum interference is a new idea in electronic
applications. Several switching devices have been proposed
wherein one can control the relative phase difference between
two interfering paths by applying electrostatic potential or
magnetic fields \cite{sols,pda1}. The possibility of transistor
action (quantum modulated transistor) in T shaped structures by
varying effective length of an open ended lead have been
explored \cite{sols,sdat}. Electron transmission in these
devices is controlled by a remote gate voltage in a region where
no classical current flows. The transmission across these
devices can be varied between 0 and 1(100$\%$ modulation), for
propagation in the fundamental transverse mode (single channel
regime). This requires that the Fermi energy should lie between
ground and the first excited transverse mode. Only in this
regime can one observe sharp and controllable variations in
transmission. However, if more than single mode propagation is
allowed, different modes with different wavelengths due to mode
mixing will produce a more complex transmission pattern and
oscillations in the total transmission are averaged out. In
fact, quantum wires with a few modes have become a reality in
the past few years.  Devices operating in the fundamental mode
promise to be much faster and will consume less power than the
conventional devices. They can also drastically reduce the size
of the electronic devices.

The conventional transistors operate in a classical diffusive
regime, and are not very sensitive to variations in material
parameters such as the dimensions or the presence of small
impurity. These devices operate by controlling the carrier
density of quasiparticles. Whereas the proposed quantum devices
are not very robust in the sense that the operational
characteristics depend very sensitively on material parameters
\cite{land1,pare}. Incorporation of a single impurity in the
mesoscopic device may change nontrivially the interference of
partial electron waves propagating through the sample and hence
the electron transmission (operational characteristics) across
the sample \cite{pare}. In such a device the actual problem of
control and reproducibility of operating threshold becomes
highly nontrivial. These devices can be exploited if we achieve
the technology that can reduce or control the phase fluctuation
to a small fraction of 2$\pi$ \cite{land1}.  On the positive
side, it should be noted that quantum devices can exhibit
multifunctional property (e.g., single stage frequency
multiplier) wherein the functions of an entire circuit can be
performed within a single element \cite{poro}.

Transport properties arising from a fundamental mode propagation
can be easily understood by a one dimensional (single channel)
modeling of the system. In such a case potential felt by an
electron can be related to the transverse width of the channel.
Naturally, modulation of the width of the conducting channel
gets related to spatial variation in the potential. In our
earlier work \cite{pda2} we have studied transport across a
metallic loop in the presence of magnetic flux. In that case
potential felt by an electron in the loop is $V \ne 0$ and in
the ideal connecting wires $V = 0$. If the Fermi energy of the
electron is less than $V$, the electron on entering the loop
propagates as evanescent modes. In such a situation, the
contribution to the conductance arises from two non-classical
effects, namely, Aharonov--Bohm effect and quantum tunneling.
For this case we have shown that, on application of a small
magnetic field, the conductance always decreases, or small field
differential magnetoconductance is always negative. This is in
contrast to the behavior in the absence of barrier, where the
small field differential magnetoconductance is negative or
positive depending on the Fermi energy and other geometric
details.  The negative differential magnetoconductance for the
case of evanescent modes can play an useful role in device
operation.  In our present treatment we have studied transport
across a modified metallic loop in the presence of a magnetic
flux. The propagation of electrons in the loop is via evanescent
modes. We have obtained an analytical expression for the
transmission coefficient and studied the effect of impurity on
the magnetoconductance of the system. We show that in the
presence of impurity, the conductance can still exhibit negative
differential magnetoconductance on the application of magnetic
field, provided the defects do not create resonant states in the
system. The change in the magnetoconductance can be made large
by intentionally incorporating impurities. This fact may play an
useful role in the device operation.

\noindent{\bf Theoretical treatment}

\noindent In our present work we have considered a system of
metallic loop connected to ideal leads as shown in Fig.1. The
geometry considered here is a modified form of geometry
considered in an earlier work \cite{pda2}. The upper arm of the
ring is of length $l_2$ and the lower arm of length $l_3$ such
that $L = l_2 + l_3$ is the circumference of the ring. The
quantum mechanical potential $V$ is positive and nonzero
(barrier) in the regions drawn as thick lines whereas zero in
the regions drawn as thin lines. The thick lines protrude into
the leads for a length $l_1$ to the left and $l_4$ to the right.
In our present problem we always take $l_1 = l_4$.  The wave
vector for the electron in the thin line region is $k =
\sqrt{E}$ where $E$ is incident Fermi energy (we have set
$\hbar$ = 2$m$ = 1). Wave vector in the thick line region is $q
= \sqrt{E-V}$. The thick line region is considered as a system
and the thin line regions are ideal leads, connecting the system
to two reservoirs on two sides. We would like to emphasize that
our one dimensional modeling of the system along with ideal
wires corresponds to a situation where electrons propagate only
in the fundamental transverse mode in a quasi--2D system. If we
have a situation in which the transverse width of the system is
much less than the ideal wires, then due to the higher zero
point energy arising from transverse confinement, fundamental
subband minima in the system will be at higher energy than the
value of a few subband minima in the ideal connecting wires.
Then a situation can arise, where several propagating modes in
the wire will have energy less than the minimum propagating
subband energy in the system. Thus the electron propagating in a
fundamental subband of the ideal wire feels a barrier to its
motion (arising solely from the mismatch in the zero point
energies) and electron tunnels across the system (due to
evanescent mode propagation) experiencing an effective potential
$V$. In our 1--D modeling in the presence of Aharonov--Bohm
magnetic flux ($\phi$), if the incident energy is less than $V$,
contribution to the transmission coefficient comes from two
non-classical effects, namely, Aharonov--Bohm effect and quantum
tunneling. We have also incorporated a delta function impurity
of strength $V_0$ at a distance $l_5$ from the sample, marked as
X in Fig.1. This will help us in understanding the role of
impurities on the magnetoconductance behavior. Transport
properties across a metallic loop for the case $V$ = 0, and
$V_0$ = 0 have been studied earlier by several workers
\cite{wash,noren,lund,das,mcle,xia,gefe}.  Following the quantum
wave guide theory on the network \cite{mod7,jaya,sadr} one can
readily calculate the transmission coefficient across the system
and is given by
\begin{equation}
T = 8k^4q^2[2 - cos(2l_2q) - cos(2l_3q) + 4 cos(\alpha) 
sin(l_2q) sin(l_3q)]/\Omega  
\end{equation}
\noindent Where $\alpha = 2\pi \phi / \phi_0, ~\phi_0 = hc/e$,
and $\phi$ is total flux piercing the loop. Expression for
$\Omega$ is too long to reproduce here, we have given the
expression for $\Omega$ separately in the appendix {\bf A}. As
expected the transmission coefficient is periodic in flux $\phi$
with period $\phi_0$ (Aharonov--Bohm oscillation) and, moreover,
$T$ is symmetric in $\phi$. The transmission coefficient $T$ is
related to two probe conductance ($G$) by the Landauer formula
$G$ = $(2e^2/h) T$ and to the dimensionless conductance g by g =
$G/\frac{2e^2}{h} = T$. Our expression (1) in the limit $V$ = 0
and $V_0$ = 0 gives the expression obtained earlier in
\cite{xia}.

\noindent{\bf Results and discussions}

\noindent We first show that design imperfections and Fermi
energy can alter the nature of output characteristics or the
conductance of the system in the presence of the propagating
modes. In order to have propagating modes we set $V$ equal to 0.
Also we set $V_0$ to 0 so that our system becomes the same as
the earlier studied system
\cite{wash,noren,lund,das,mcle,xia,gefe} (this ensures that q
$\rightarrow$ k in equation (1)).  Design imperfections may lead
to variations in the arm lengths of the loop. In fig.2 we have
plotted the dimensionless conductance g ($\equiv \mid T \mid^2$)
versus dimensionless magnetic flux $\alpha$ for $l_2\over L$ =
0.2, $l_3\over L$ = 0.8, for different values of $kL$
(dimensionless Fermi wave vector). Although all the curves
oscillate with a period of $2\pi$, as expected
\cite{wash,noren,lund,das,mcle,xia,gefe}, these curves are,
however, completely diverse in nature. The solid curve is
plotted for $kL = 0.2$. It shows that the small field
differential magnetoconductance is negative. The dashed curve
and the dotted curve are plotted for $kL =2.0$ and $kL =3.5$,
respectively. Both of them show more oscillations than the solid
curve. The small field differential magnetoconductance is
positive for both.  However, for the dashed curve the absolute
minima in the conductance occurs at $\alpha = \pi$ whereas for
the dotted curve the absolute minima in the conductance occur at
$\alpha = 0$. Similar diversities can be seen if we fix the
Fermi energy and vary other parameters such as $l_2 \over L$ and
$l_3 \over L$. One can readily check this from the fact that
$kl_2$, $kl_3$, etc., always occur as a single variable in the
expression for the dimensionless conductance g. A single
impurity inside the ring or in the lead can also drastically
alter the output characteristics \cite{pare}. In fig.3 we have
shown the sensitivity in output characteristics in presence of
propagating modes ({\em i.e.}, $V$ = 0 and q $\rightarrow k$ in
equation (1)), due to a single defect in the lead and design
imperfections. In this figure we have plotted the conductance
versus $kL$ for different cases in the absence of magnetic
field. In the case of the solid curve $\frac{l_2}{L} = 0.5$,
$\frac{l_3}{L} = 0.5$ and $V_0L$ = 0. In the case of the dashed
curve $\frac{l_2}{L} = 0.75$, $\frac{l_3}{L} = 0.25$ and $V_0L =
0$ whereas the dotted curve is for $\frac{l_2}{L} = 0.5$,
$\frac{l_3}{L} = 0.5$, $V_0L = 1.0$ and $\frac{l_5}{L} = 1.0$.
In all these cases we have set $\alpha$ = 0. From this figure
one can readily notice the sensitive dependence of the output
characteristic on material parameters. With such sensitivity of
output characteristics with respect to the variation in the
Fermi energy, the length parameters and the magnetic flux (fig.2
and fig.3) it is difficult to ensure stability in the device
performance. This problem does not arise in the case of
evanescent modes in the system. In fig.4, we have plotted the
conductance g due to evanescent modes versus magnetic field
$\alpha$ for three different values of Fermi energy. We have
fixed other parameter values as $\frac{l_{1}}{L} = 0.1$,
$\frac{l_{2}}{L}$ = 0.2, $\frac{l_{3}}{L}$ = 0.8, $VL$ = 16 and
$V_0$ = 0.  The dotted curve, the dashed curve and the solid
curve corresponds to the $kL$ values 3.5, 3.0 and 2.0,
respectively.  To have evanescent modes in the ring it is
necessary to take a sufficiently large $V$ such that the
incident Fermi energy $E < V$. $V_0$ is still set to zero.
Transmission occurs due to quantum mechanical tunneling and it
is to be calculated by analytic continuation which means in the
equation (1) we have to put q $\rightarrow$ iQ, where Q=$\sqrt{V
- k^2}$.  We find that for all the three Fermi energies
considered in fig.4 the conductance versus $\alpha$ plots are
similar. All of them initially show negative differential
magnetoconductance with the absolute minima occurring at $\alpha
= \pi$. The physical reason for this is that for evanescent
modes it is always the first harmonic in the Fourier expansion
(in $\alpha$) of transmission that dominates over all the others
in determining the conductance \cite{pda2}, which is not the
case for the propagating modes.  In the case of propagating
modes for a particular configuration or for a particular Fermi
energy, on the other hand, any one of the infinite number of
harmonics in the Fourier expansion of transmission coefficient
can give large contributions and hence the output
characteristics changes drastically with change in the
configuration or the Fermi energy.  Such a systematic nature of
output characteristics for evanescent modes will help devising
robust switches. However the price we pay is loss in the
sensitivity of conductance to change of $\alpha$. This is
because evanescent modes are not so sensitive to changes in the
boundary conditions induced by the change in the flux $\alpha$.

In this work we further investigate the role of impurities in
the connecting leads  in modifying the above mentioned features.
By definition electrons will always be in propagating modes in
the connecting leads. Impurities inside the ring are not
expected to produce any drastic changes if we have evanescent
modes in the ring. This is due to the fact that, having
evanescent modes in the ring the impurities cannot cause any
additional resonance.  Any additional phase change due to
impurity scattering is almost equivalent to changing the
effective arm lengths of the two loops in the ring. Thus to
expect nontrivial effects due to impurities in the presence of
evanescent modes in the system, impurities must be located in
the connecting leads where electrons travel as propagating
modes. To study this we have incorporated a single delta
function potential of strength $V_0$ at a distance $l_5$ from
the loop at the point $X$ shown in fig.1.  We choose the typical
values of parameters such that $\frac{l_1}{L} = 0.1$, $l_2 \over
L$ =0.5, $l_3 \over L$ = 0.5, $l_5 \over L$ = 1.0, $VL$ = 5.0.
Note that $l_2 = l_3$, { \em i.e.} we are considering a
symmetric loop.  For such a system we have plotted in fig.5 the
conductance g versus incident Fermi wave vector $kL$ in the
energy interval where we have only evanescent modes in the ring.
Choosing $V_0 L$ = 0 we have plotted the conductance g versus kL
for two values of $\alpha$ {\em i.e.} $\alpha = 0$ (dotted
curve) and $\alpha = 2$ (dot--dashed curve).  Also for $V_0 L$ =
1 we have plotted the conductance g versus kL for the same two
values of $\alpha$ {\em i.e.} $\alpha = 0$ (solid curve) and
$\alpha = 2$ (dashed curve).  We find by comparing the dashed
and solid curves that the magnetic field decreases the
conductance at all values of $kL$ signifying that the
magnetoconductance of the system is still negative inspite of
the impurity in one of the leads.  Qualitatively there is no
change compared to the situation when $V_{0}L$ = 0.  This
feature remains unchanged as one increases the impurity strength
$V_0$. However, for large $V_0$ one can have a resonance in the
energy range 0 to $\sqrt V$ due to multiple scattering in the
region of length $l_5$. This is shown in fig.6 where we have
plotted conductance versus $kL$ for the same two values of
$\alpha$ and the same set of parameters as in fig.5 {\em i.e.}
$\alpha$ = 0 (solid curve) and $\alpha$ = 2 (dashed curve) for
$\frac{l_1}{L} = 0.1$, $\frac{l_2}{L} = 0.5$, $\frac{l_3}{L} =
0.5$, $\frac{l_5}{L} = 1.0$ and $VL = 5.0$. Only the strength of
the delta potential has been increased to $V_0L$ = 3 as compared
to the situation in fig.5.  One can clearly see a well defined
conductance peak or a resonance in the absence of magnetic flux
($\alpha = 0$). For nonzero $\alpha$ the scattering strength of
the ring is much higher than that of the impurity (differential
magnetoconductance of isolated ring being negative definite,
magnetic field increases the scattering strength or decreases
the transmission coefficient of the loop) and this rules out the
possibility of a sharp resonance.  Superposed in this graph we
have also shown the $V_{0}L$ = 0 situation for the same two
values of $\alpha$ {\em i.e.} $\alpha = 0$ (dotted curve) and
$\alpha = 2$ (dot--dashed curve) for the same values of the
other parameters. A comparison of the two sets of graphs (each
for $V_0L \ne 0$ and $V_0L = 0$) shows that the resonance not
only increases the conductance but also increases the
sensitivity of the conductance to change in magnetic field or
$\alpha$. At a Fermi energy marked as $k_{1}L$ in Fig.6, by
increasing the magnetic field or $\alpha$ we can decrease the
conductance much more when $V_{0}L$ = 3 than when $V_{0}L$ = 0.
This is true for any kL in general.  To show this, in fig.7. we
have fixed the incident energy such that $kL$ = 1.4 (solid
line), $kL$ = 1.6 (dashed line) and $kL$ = 1.8 (dotted line) for
$\frac{l_1}{L} = 0.1$, $l_2 \over L$ = 0.5, $l_3 \over L$ = 0.5,
$l_5 \over L$ = 1.0, VL=5.0 and plotted ($g(\alpha= 0) -
g(\alpha =2)$) (this quantity can be taken as a measure of the
magnitude of differential magnetoconductance or a measure of the
flux sensitivity of the conductance) versus $V_0L$ {\em i.e.}
the strength of the delta function potential. We find that
initially the flux sensitivity of the system increases with the
impurity strength, passes through a peak, and asymptotically
decreases. For $V_0L= 0$ there can be no resonant transmission.
Initially as $V_0$ increases we approach the resonance between
the loop and the delta potential. This resonance not only
increases the conductance but also increases the sensitivity of
conductance to twisting of boundary condition by the  magnetic
field and hence the differential magnetoconductance increases in
magnitude while being negative all the time. However, for very
large value of $V_0 L$ we move far away from resonance condition
and then due to enhanced scattering by the delta potential the
flux sensitivity of the conductance or the differential
magnetoconductance decreases.  In fig.7 as ($g(\alpha =0) -
g(\alpha =2)$) is positive over the whole range of $V_0L$ the
magnetoconductance is negative at these Fermi energies for the
whole range of $V_0L$ shown.

Notably, in fig.6, inspite of the resonance the
magnetoconductance is negative. However if the impurity strength
is slightly higher than the scattering strength of the isolated
ring then there can be sharp resonances in the absence of
magnetic field ($\alpha$ = 0) as well as in the presence of
magnetic field ($\alpha$ = 2). The reason is that for both
$\alpha = 0$ and $\alpha \ne 0$ the scattering strength of the
loop is comparable (in one case it is slightly smaller and in
the other case it is slightly higher) to that of the delta
potential.  Then the resonant conductance at $\alpha$ = 2 can
exceed the nonresonant conductance at $\alpha$ = 0 implying
positive differential magnetoconductance in a small range of
Fermi energy. This is shown in fig.8, where we have plotted
dimensionless conductance g versus kL for $\alpha = 0$ (solid
line) and $\alpha = 2$ (dashed line) for $\frac{l_1}{L} = 0.1$,
$\frac{l_2}{L} = 0.5, \frac{l_3}{L} = 0.5, \frac{l_5}{L} = 1.0,
VL = 5$ and $V_{0}L$ = 5.5. In a small energy range { \em i.e.}
$kL$ = 2.1 to 2.25 the $\alpha$ = 2 (dashed) curve goes above
the $\alpha$ = 0 (solid) curve signifying positive
magnetoconductance in this energy window although we have only
evanescent modes in the ring.  Hence the scattering strength of
the impurity should lie above a critical value which is higher
than the scattering strength of the loop. Otherwise the magnetic
field will increase the scattering strength of the loop (the
magnetoconductance of the isolated loop being negative definite)
thus detuning the strength of the two scatterers and making
sharp resonance for nonzero $\alpha$ impossible, within the
relevant energy window. When a sharp resonance appears the
output characteristics become very sensitive to the material
parameters. For small field, the change in the
magnetoconductance can become either negative or positive with
the absolute minima at $\alpha = 0$ as well as at $\alpha = \pi$
depending on the exact choice of Fermi energy. This behavior is
similar to that in the presence of propagating modes along the
ring, {\em e.g.} see fig.2.

These features of the output characteristics in the presence of
defect, remain unchanged even if there are design imperfections
like unequal arm lengths.  Again, positive differential
magnetoconductance occurs only when the resonant conductance at
$\alpha \neq$ 0 exceeds the nonresonant conductance at $\alpha
=0$. In this case as one of the arms is much shorter than the
other it shunts most of the current and reduces the scattering
strength of the loop. Hence resonances can occur for smaller
values of the delta potential strength than those in the case of
fig.8. If $\frac{l_2}{L}$ =0.2 and $\frac{l_3}{L}$ = 0.8 then
$V_0L$ = 5 can give rise to appreciable positive
magnetoconductance in a small energy range for the same values
of the other parameters as in Fig.8.

It is evident from fig.8 that even if the delta potential
strength is strong enough to produce resonance, it reduces the
energy window where we have negative differential
magnetoconductance, by a small amount. The range of Fermi wave
vector in which we have only evanescent modes in the sample is k
= 0 to k = $\sqrt{V}$. The resonance condition being $k l_5 =
2\pi$, we cannot have a resonance in the relevant energy range
unless $l_5 \ge \frac{2\pi}{\sqrt{V}}$.  So for $l_5 \le \frac{2
\pi}{\sqrt{V}}$ we can never have positive magnetoconductance in
the presence of evanescent modes.

However if $l_5 >> \frac{2\pi}{\sqrt{V}}$ then one can have many
resonances in the energy interval where we can have only
evanescent modes in the loop. And whenever there is a resonance,
there is a possibility of obtaining positive differential
magnetoconductance. This is shown in fig.9, where we have
plotted conductance for two values of $\alpha$ {\em i.e.}
$\alpha$ = 0 (solid line) and $\alpha$ = 2 (dotted line) versus
incident energy $kL$ for $\frac{l_1}{L} = 0.1$, $l_2 \over L$ =
0.2, $l_3 \over L$ = 0.8, $l_5 \over L$ = 5.0, $VL = 5.0$  and
$V_0L$ = 5.0.  We find that in the energy window where we have
only evanescent modes in the loop there can be three resonances.
Near each of these resonances in a very narrow energy window the
dotted curve exceeds the solid curve signifying that
differential magnetoconductance is positive in these regions. At
other energies change in magnetoconductance is always negative.

In conclusion, we have investigated transmission across a loop
in the presence of magnetic field and an impurity. In our case
electronic wave travels as evanescent waves throughout the
circumference of the loop. In such a situation we have shown
that initial change in the magnetoconductance is negative even
in the presence of impurities, provided impurities do not create
resonant states in the system. For small fields the change in
magnetoconductance can be made large by intentionally
incorporating impurity. This fact can be used for an operation
of a quantum switch. Where the on and the off states can
correspond to transmission in the absence or in the presence of
magnetic field, respectively {\em i.e.}, the on state has always
a larger conductance than the off state.  The magnitude of
negative differential magnetoconductance may also be enhanced in
a multichannel situation. In this case, one can populate
electrons in connecting leads in many lower subband channels
(multichannels) corresponding to different transverse quantum
numbers, such that the Fermi energy lies below the lowest
subband of the loop. This can be achieved by making the width of
the loop much smaller than the width of the connecting leads,
{\em i.e.} the quantum zero point energy in the loop will be
much higher than several subband energies in the connecting
leads. Then all these subbands in the leads will contribute to
the conductance through evanescent modes.

\newpage

\noindent{\large{\bf Appendix A}}

\noindent{\bf An expression for $\Omega$}

\noindent $\Omega$ = ($C^2$ + $D^2$)

\noindent $C$ =  cos(k$l_5$) cos($\alpha$) 
{\bf (} 2$k^2$$V_0$  cos(2$l_1$q) -- 2$k^2$$V_0$ 
-- 4$k^2$q sin(2$l_1$q) {\bf )}
+ cos($\alpha$) sin(k$l_5$) {\bf (} 2$k^3$ -- 2k$q^2$ 
-- 2$k^3$ cos(2$l_1$q) -- 2k$q^2$ cos(2$l_1$q) 
-- 2kq$V_0$ sin(2$l_1$q) {\bf )}
+  cos($l_2$q) cos($l_3$q) {\bf (} 2$k^2$$V_0$ cos(k$l_5$)
-- 2$k^2$$V_0$ cos(k$l_5$) cos(2$l_1$q) -- 2$k^3$ sin(k$l_5$)
+ 2k$q^2$ sin(k$l_5$) + 2$k^3$ cos(2$l_1$q) sin(k$l_5$)
+ 2k$q^2$ cos(2$l_1$q) sin(k$l_5$)
+ 4$k^2$q cos(k$l_5$) sin(2$l_1$q)
+ 2kq$V_0$ sin(k$l_5$) sin(2$l_1$q) {\bf )}
+ {\bf (} 4$k^2$q cos(k$l_5$) cos(2$l_1$q) 
+ 2kq$V_0$ cos(2$l_1$q) sin(k$l_5$) 
+ 2$k^2$$V_0$ cos(k$l_5$) sin(2$l_1$q) 
-- 2k$q^2$ sin(k$l_5$) sin(2$l_1$q) 
-- 2$k^3$ sin(k$l_5$) sin(2$l_1$q) {\bf )}
{\bf (} cos($l_3$q) sin($l_2$q) + cos($l_2$q) sin($l_3$q){\bf )}
-- sin($l_2$q) sin($l_3$q) {\bf (} $k^2$$V_0$ cos(k$l_5$) 
+ 3$k^2$$V_0$ cos(k$l_5$) cos(2$l_1$q) 
+ $k^3$ sin(k$l_5$) -- k$q^2$ sin(k$l_5$) 
-- 3$k^3$ cos(2$l_1$q) sin(k$l_5$) 
-- 3k$q^2$ cos(2$l_1$q) sin(k$l_5$) 
-- 6$k^2$q cos(k$l_5$) sin(2$l_1$q) 
-- 3kq$V_0$ sin(k$l_5$) sin(2$l_1$q) {\bf )} 
\vskip .5cm

\noindent $D$ = cos(k$l_5$) cos($\alpha$) 
{\bf (}2k$q^2$ -- 2$k^3$ 
+ 2$k^3$ cos(2$l_1$q) + 2k$q^2$ cos(2$l_1$q) + 2kq$V_0$
sin(2$l_1$q) {\bf )}
+ cos($l_2$q) cos($l_3$q) {\bf (}2$k^3$ cos(k$l_5$) 
-- 2k$q^2$ cos(k$l_5$) -- 2$k^3$ cos(k$l_5$) cos(2$l_1$q) 
-- 2k$q^2$ cos(k$l_5$) cos(2$l_1$q) -- 2$q^2$$V_0$ sin(k$l_5$)
-- 2$q^2$$V_0$ cos(2$l_1$q) sin(k$l_5$) -- 2kq$V_0$ cos(k$l_5$)
sin(2$l_1$q) 
+ 4$k^2$q sin(k$l_5$) sin(2$l_1$q) {\bf)} +
 cos($\alpha$) sin(k$l_5$) {\bf (} 2$q^2$$V_0$ + 2$q^2$$V_0$
cos(2$l_1$q) 
-- 4$k^2$q sin(2$l_1$q) {\bf )}
+ {\bf (} 4$k^2$q cos(2$l_1$q) sin(k$l_5$) 
-- 2kq$V_0$ cos(k$l_5$) cos(2$l_1$q) + 2$k^3$ cos(k$l_5$) sin(2$l_1$q)
+ 2k$q^2$ cos(k$l_5$) sin(2$l_1$q)
+ 2$q^2$$V_0$ sin(k$l_5$) sin(2$l_1$q) {\bf )} {\bf (}cos($l_3$q)
sin($l_2$q) + cos($l_2$q) sin($l_3$q) {\bf)}
-- sin($l_2$q) sin($l_3$q) {\bf (} $k^3$ cos(k$l_5$) 
+k$q^2$ cos(k$l_5$) + 3$k^3$ cos(k$l_5$) cos(2$l_1$q) 
+ 3k$q^2$ cos(k$l_5$) cos(2$l_1$q) + $q^2$$V_0$ sin(k$l_5$) 
+ 3$q^2$$V_0$ cos(2$l_1$q) sin(k$l_5$) 
+ 3kq$V_0$ cos(k$l_5$) sin(2$l_1$q) 
-- 6$k^2$q sin(k $l_5$) sin(2$l_1$q) {\bf )}

\newpage

\noindent{\large{\bf Figure Captions}}

\noindent {\bf Fig. 1.} A normal metal ring connected to two
electron reservoirs by ideal leads. Quantum mechanical potential
in the sample is V. At site X there is a delta function impurity
potential of strength $V_0$. The magnetic flux $\phi$ pierces
through the ring. The different lengths $l$ are marked in the
figure.

\noindent {\bf Fig. 2.} Plot of conductance g versus $\alpha$
for kL = 0.2 (solid curve), kL = 2.0 (dashed curve) and kL = 3.5
(dotted curve). $\frac{l_1}{L}$ = $\frac{l_4}{L}$ =
$\frac{l_5}{L}$ = 0, $\frac{l_2}{L}$ = 0.2, $\frac{l_3}{L}$ =
0.8, VL = 0 and $V_0L = 0$ for all the cases.

\noindent {\bf Fig. 3.} Plot of conductance g versus kL with
and without impurity when electrons travel in propagating modes.
$\frac{l_2}{L}$ = $\frac{l_3}{L}$ = 0.5, $\frac{l_5}{L}$ = 0,
$V_0$L = 0 (solid curve), $\frac{l_2}{L}$ = $\frac{l_3}{L}$ =
0.5, $\frac{l_5}{L}$ = 1.0, $V_0$L = 1.0 (dotted curve),
$\frac{l_2}{L}$ = 0.75, $\frac{l_3}{L}$ = 0.25, $\frac{l_5}{L}$
= 0, $V_0$ = 0 (dashed curve) and $\frac{l_1}{L}$ =
$\frac{l_4}{L}$ = 0, VL = 0 for all the cases.

\noindent {\bf Fig. 4.} Plot of conductance versus $\alpha$ for
different values of kL when electron in evanescent modes. kL =
2.0 (solid curve), kL = 3.0 (dashed curve), kL = 3.5 (dotted
curve).  $\frac{l_1}{L}$ = $\frac{l_4}{L}$ = 0.1,
$\frac{l_2}{L}$ = 0.2, $\frac{l_3}{L} = 0.8$, $\frac{l_5}{L}$ =
1.0, VL = 16, $V_0L = 0$ for all the curves.

\noindent {\bf Fig. 5.} Plot of conductance g versus kL in
presence and absence of both impurity ($V_0$) and magnetic flux
($\alpha$).  $V_0L = 0$ and $\alpha$ = 0 (dotted curve), $V_0L =
0$ and $\alpha$ = 2 (dot-dashed curve), $V_0L = 1.0$ and
$\alpha$ = 0 (solid curve), $V_0L = 1.0$ and $\alpha$ = 2
(dashed curve).  $\frac{l_1}{L}$ = $\frac{l_4}{L}$ = 0.1,
$\frac{l_2}{L}$ = $\frac {l_3}{L} = 0.5$, $\frac{l_5}{L}$ = 1.0,
VL = 5.0 for all the curves.

\noindent {\bf Fig. 6.} Plot of conductance g versus kL in
presence and absence of both impurity and magnetic flux when
electrons travel in evanescent modes. $V_0L = 0$ and $\alpha$ =
0 (dotted curve), $V_0L = 0$ and $\alpha$ = 2 (dot-dashed
curve), $V_0L = 3.0$ and $\alpha$ = 0 (solid curve), $V_0L =
3.0$ and $\alpha$ = 2 (dashed curve). All other parameters are
same as in fig.5.

\noindent {\bf Fig. 7.} Plot of the difference between the
conductances (g(0)--g(2)) in the absence ($\alpha$ = 0) and in
the presence ($\alpha$ = 2) of magnetic flux  versus impurity
($V_0L$) when electrons travel in evanescent modes. kL = 1.4
(solid curve), kL = 1.6 (dashed curve) and kL = 1.8 (dotted
curve). All other parameters are same as in fig.5.
\newpage
\noindent {\bf Fig. 8.} Plot of conductance g versus kL in the
presence and in the absence of magnetic flux with impurity.
$V_0L = 5.5$ and $\alpha$ = 0 (solid curve), $V_0L = 5.5$ and
$\alpha$ = 2 (dashed curve). All other parameters are same as in
fig.5.

\noindent {\bf Fig. 9.} Plot of conductance g versus kL in
presence or absence of magnetic flux with impurity.
$\frac{l_1}{L}$ = $\frac{l_4}{L}$ = 0.1, $\frac{l_2}{L}$ = 0.2,
$\frac{l_3}{L}$ = 0.8, $\frac{l_5}{L}$ = 5.0, VL = 5.0, $V_0L$ =
5.0, $\alpha$ = 0 (solid curve) and $\alpha$ = 2 (dotted curve).


\newpage

\begin{thebibliography}{99}

\bibitem{kirk} See e,g,. Physics and Technology of Submicron
Structures, ed. H.  Heinrich, G. Bauer and F. Kuchar (Springer,
New-York 1988).  Nanostructure Physics and Fabrication, ed. M.
A. Reed and W. P. Kirk (Academic Press, New York 1990).
\bibitem{been1} C. W. J. Beenakker and H. van Houten, in Solid
State Physics: Semiconductor Heterostructures and
Nanostructures, Vol 44, ed. H.  Ehrenreich and D. Turnbull,
Academic Press(1991).
\bibitem{kram} Quantum Coherence in Mesoscopic Systems, ed. B.
Kramer, NATO ASI Series B, Vol 254, Plenum(1991).
\bibitem{alts1} Mesoscopic Phenomenon in Solids, ed. B. L.
Altshular, P. A. Lee and R. A. Webb, North Holland(1991).
\bibitem{bout} F. A. Bout, Phys. Rep. {\bf 234} 73 (1993).
\bibitem{sols} F. Sols, M. Macucci, V. Ravoili and K. Hess.
Appl. Phys. Lett.  {\bf 54}, 350(1990); J. Appl. Phys. {\bf 66},
3892(1989).
\bibitem{pda1} P. Singha Deo and A. M. Jayannavar, Phys. Rev. B
{\bf 50} 11629 (1994) and references therein.
\bibitem{sdat} S. Datta, Superlatt. Microstruct. {\bf 6} 83 (1989).
\bibitem{land1} R. Landauer, Physics Today, {\bf 42} (No. 10), 119(1989).
\bibitem{pare} A. M. Jayannavar, P. Singha Deo and T. P. Pareek,
Cluster and Nano-structured materials, ed. P. Jena and S. N.
Behera, Nova Science Publication, New york (1996) in print.
\bibitem{poro} S. Subramaniam, S. Bandyopadhyay and W. Porod, J.
Appl.  Phys. {\bf 68}, 4861(1990).
\bibitem{pda2} P.  Singha.  Deo.  and A.  M.  Jayannavar, Mod.
Phys. Lett. B8, 301 (1994).
\bibitem{wash} S. Washburn and R. A. Webb, Adv. Phys. {\bf 35}, 75(1986).
\bibitem{noren} S. Datta, M. R. Melloch, S. Bandyopadhyay, R.
Noren, M.  Vaziri, M. Miller and R. Reifenberger, Phys. Rev.
Lett. {\bf 55} 2344 (1986).
\bibitem{lund} S. Datta, M. R. Melloch, S. Bandyopadhyay and M.
S. Lundstrom, Appl. Phys. Lett. {\bf 48}, 487(1986).
\bibitem{das} S. Datta and B. Das, Appl. Phys. Lett. {\bf 56}, 665 (1990).
\bibitem{mcle} S. Datta and M. J. McLennan. Rep. Prog. Phys.
{\bf 53}, 1003(1990) and references therein.
\bibitem{xia} J. Xia, Phys. Rev. B {\bf 45}, 3593 (1992).
\bibitem{gefe} Y. Gefen, Y. Imry and M. Ya Azbel, Phys. Rev.
Lett. {\bf 52}, 139(1984).
\bibitem{mod7} P.  Singha Deo and A.  M.  Jayannavar, Mod. Phys.
Lett.  B {\bf 7}, 1045(1993)
\bibitem{jaya} A.  M.  Jayannavar and P.  Singha.  Deo, Phys.
Rev.  B {\bf 49},13685 (1994).
\bibitem{sadr} A. F. Sadreeb, V. A. Vid$'$manov, Int. Journ. of
Mod. Phys.  B {\bf 9} 2719 (1995).
\end{thebibliography}
\end{document}